\documentclass[apj]{emulateapj}
\slugcomment{Submitted to AJ; August 19, 2007}

\newcommand{\lya}{Ly$\alpha$}
\newcommand{\nv}{N\,{\sc v}}
\newcommand{\siii}{Si\,{\sc ii}}
\newcommand{\oi}{O\,{\sc i}}
\newcommand{\lyb}{Ly$\beta$}
\newcommand{\ovi}{O\,{\sc vi}}
\newcommand{\civ}{C\,{\sc iv}}
\newcommand{\civd}{C\,{\sc iv} $\lambda\lambda$1548,1550}
\newcommand{\mgii}{Mg\,{\sc ii}}

\begin{document}

\title{A SURVEY OF {$z\sim6$} QUASARS IN THE SDSS DEEP STRIPE: I.
A FLUX-LIMITED SAMPLE AT {$z_{AB}<21$}}\thanks{
Based on observations obtained with the Sloan Digital Sky Survey, which is
owned and operated by the Astrophysical Research Consortium; the MMT 
Observatory, a joint facility of the University of Arizona and the Smithsonian 
Institution; the 6.5 meter Magellan Telescopes located at Las Campanas 
Observatory, Chile; the W.M. Keck Observatory, which is operated as a 
scientific partnership among the California Institute of Technology, the 
University of California and the National Aeronautics and Space 
Administration, and was made possible by the generous financial support of 
the W.M. Keck Foundation.}

\author{Linhua Jiang\altaffilmark{1}, Xiaohui Fan\altaffilmark{1}, 
  James Annis\altaffilmark{2}, Robert H. Becker\altaffilmark{3,4},
  Richard L. White\altaffilmark{5},
  Kuenley Chiu\altaffilmark{6}, Huan Lin\altaffilmark{2}, 
  Robert H. Lupton\altaffilmark{7}, Gordon T. Richards\altaffilmark{8},
  Michael A. Strauss\altaffilmark{7}, Sebastian Jester\altaffilmark{9},
  and Donald P. Schneider\altaffilmark{10}}
\altaffiltext{1}{Steward Observatory, University of Arizona,
  933 North Cherry Avenue, Tucson, AZ 85721}
\altaffiltext{2}{Fermi National Accelerator Laboratory, P.O. Box 500, 
  Batavia, IL 60510}
\altaffiltext{3}{Physics Department, University of California, Davis, CA 95616}
\altaffiltext{4}{Institute of Geophysics and Planetary Physics, Lawrence 
  Livermore National Laboratory, Livermore, CA 94550}
\altaffiltext{5}{Space Telescope Science Institute, 3700 San Martin Drive, 
  Baltimore, MD 21218}
\altaffiltext{6}{School of Physics, University of Exeter, Stocker Road, 
  Exeter EX4 4QL, UK}
\altaffiltext{7}{Department of Astrophysical Sciences, Princeton University,
  Princeton, NJ 08544}
\altaffiltext{8}{Department of Physics, Drexel University, 3141 Chestnut 
  Street, Philadelphia, PA 19104}
\altaffiltext{9}{Max-Planck-Institut f\"{u}r Astronomie, K\"{o}nigstuhl 17,
  D-69117 Heidelberg, Germany}
\altaffiltext{10}{Department of Astronomy and Astrophysics, Pennsylvania
  State University, 525 Davey Laboratory, University Park, PA 16802}

\begin{abstract}

We present the discovery of five quasars at $z\sim6$ selected from 260 deg$^2$ 
of the Sloan Digital Sky Survey (SDSS) southern survey, a deep imaging survey 
obtained by repeatedly scanning a stripe along the Celestial Equator. 
The five quasars with $20<z_{AB}<21$ are 1--2 magnitudes fainter than the 
luminous $z\sim6$ quasars discovered in the SDSS main survey. One of them was 
independently discovered by the UKIRT Infrared Deep Sky Survey. These quasars, 
combined with another $z\sim6$ quasar known in this region, make a complete 
flux-limited quasar sample at $z_{AB}<21$. The sample spans the redshift range 
$5.85\le z\le 6.12$ and the luminosity range $-26.5\le M_{1450}\le -25.4$
(H$_0=70$ km s$^{-1}$ Mpc$^{-1}$, $\Omega_{m}=0.3$, and 
$\Omega_{\Lambda}=0.7$). 
We use the $1/V_{a}$ method to determine that the comoving quasar spatial 
density at $\langle z\rangle=6.0$ and $\langle M_{1450}\rangle=-25.8$
is $(5.0\pm2.1)\times10^{-9}$ Mpc$^{-3}$ mag$^{-1}$. We model the bright-end
quasar luminosity function (QLF) at $z\sim6$ as a power law
$\Phi(L_{1450})\propto L_{1450}^{\beta}$. The slope $\beta$ calculated from a 
combination of our sample and the luminous SDSS quasar sample is $-3.1\pm0.4$, 
significantly steeper than the slope of the QLF at $z\sim4$. Based on the 
derived QLF, we find that the quasar/AGN population cannot provide enough 
photons to ionize the intergalactic medium (IGM) at $z\sim6$ unless the IGM is 
very homogeneous and the luminosity ($L_{1450}^\ast$) at which the QLF 
power law breaks is very low.

\end{abstract}


\section{INTRODUCTION}

High-redshift quasars are among the most luminous objects known and provide 
direct probes of the distant universe when the first generation of galaxies 
and quasars formed. In recent years, over twenty $z\sim6$ quasars with 
$z_{AB}\leq20$ have been discovered \citep[e.g.][]{fan00,fan01a,fan03,fan04,
fan06a,got06}. These luminous quasars are essential for
understanding the accretion history of black holes (BHs), galaxy formation, 
and chemical evolution at very early epochs. They harbor supermassive BHs with 
masses higher than $10^9\ M_{\sun}$ and emit near the Eddington limit 
\citep[e.g.][]{bar03,ves04,jia06a,kur07}, revealing the rapid growth of 
central BHs at high redshift. Their emission lines show solar or supersolar 
metallicity in the broad line regions, indicating that there was 
vigorous star formation and element enrichment in the first gigayear of cosmic 
time \citep[e.g.][]{bar03,mai03,jia07,kur07}. Their absorption spectra show 
that the intergalactic medium (IGM) at $z\sim6$ is close to the reionization 
epoch \citep[e.g.][]{bec01,djo01,fan06b,fan06c}.

The majority of the currently known $z\sim6$ quasars were discovered from 
$\sim8000$ deg$^2$ of imaging data of the Sloan Digital Sky Survey 
\citep[SDSS;][]{yor00}. They were selected as $i$-dropout objects using 
optical colors. Several other high-redshift quasars were discovered based on 
their infrared or radio emission. For example, \citet{coo06} discovered one 
quasar at $z=5.85$ in the NOAO Deep Wide-Field Survey \citep[NDWFS;][]{jan99} 
Bootes Field using the AGN and Galaxy Evolution Survey (AGES) spectroscopic 
observations. The quasar was selected from a $Spitzer$ mid-infrared quasar 
sample and has a $z_{AB}$ magnitude of 20.68 and an optical luminosity of 
$M_B=-26.52$. By matching the FLAMINGOS Extragalactic Survey IR survey 
\citep{els06} data to the Faint Images of the Radio Sky at Twenty-cm 
\citep[FIRST;][]{bec95} data, \citet{mcg06} discovered a radio-loud quasar at 
$z=6.12$ in 4 deg$^2$ of the NDWFS region. This quasar is a broad absorption 
line (BAL) quasar with an optical luminosity of $M_B=-26.9$, comparable to 
the luminous SDSS quasars at $z\sim6$.

Despite the high-redshift quasar surveys mentioned above, very little is known 
about faint quasars ($z_{AB}>20$) at $z\sim6$. The SDSS main survey only 
probes the most luminous quasars, and with a density of 1/470 deg$^2$ 
\citep{fan06a}. The \citet{coo06} 
quasar at $z=5.85$ was $z_{AB}>20$, but the sample contains a single object
and is selected from an area of less than 10 deg$^2$. \citet{mah05} found a 
very faint quasar with $z_{AB}=23.0$ at $z=5.70$ in a 2.5 deg$^2$ field 
around the luminous quasar 
SDSS J114816.64+525150.3\footnote{The naming convention for SDSS sources 
is SDSS JHHMMSS.SS$\pm$DDMMSS.S, and the positions are expressed in J2000.0
coordinates. We use SDSS JHHMM$\pm$DDMM for brevity.} at $z=6.42$.
\citet{wil05} imaged a 3.83 deg$^2$ region down to $z_{AB}=23.35$ in the first 
results of the Canada-France High-redshift Quasar Survey (CFHQS) and did not 
find any quasars at $z>5.7$. In these surveys both the quasar samples 
and the survey 
areas are very small, thus they do not provide a good statistical study of 
high-redshift quasars at $z_{AB}>20$. Recently, \citet{wil07} discovered four 
quasars at $z>6$ from about 400 deg$^2$ of the CFHQS, including the most 
distant known quasar at $z=6.43$. Three of these quasars 
have $z_{AB}$ magnitudes 
fainter than 21. Since their follow-up observations are not yet complete, they 
did not determine the spatial density of these quasars.

Finding faint quasars at $z\sim6$ is important for studying the
evolution of the quasar population and quasars' impact on their
environments.  \citet{fan04} obtained the bright-end quasar luminosity
function (QLF) at $z\sim6$, but the slope, $-3.2\pm0.7$, was very
uncertain due to the small luminosity range of the
sample. \citet{ric04} put a broad constraint on the bright-end slope
of $>-4.63$ ($3\sigma$) from the absence of lenses in four quasars at
$z\sim6$.  \citet{sm07} have considered the implications of all
existing $z\sim6$ quasar observations, including deep X-ray surveys,
for the faint end of the high-redshift QLF. Based predominantly on the
X-ray surveys, they argue that there is a flattening of the QLF at
$M_{1450} \ga -24.67$.

With the discovery of faint high-redshift quasars, the QLF can be well
determined. The QLF at $z\sim6$ is important to understand BH growth
at early epochs \citep[e.g.][]{vol06,wyi06}. While bright quasars at
high redshift have central BH masses between $10^9$ and
$10^{10}\ M_{\sun}$, fainter quasars with $z_{AB}>20$ are expected to
harbor BHs with masses of a few times 10$^8\ M_{\sun}$ or below
\citep[e.g.][]{kur07}, which may be associated with galaxies of lower
masses. The QLF also enables us to determine the quasar contribution
to the UV background at $z\sim6$. Detection of complete Gunn-Peterson
troughs \citep{gun65} among the highest-redshift quasars indicates a
rapid increase of the IGM neutral fraction at $z\sim6$, and suggests
that we have reached the end of the reionization epoch
\citep[e.g.][]{bec01,djo01,fan06c}.  It is unclear what individual
contributions of galaxies and quasars to the reionization
are. Although there is evidence showing that quasars are probably not
the main contributor to reionization \citep[e.g.][]{sal07,srb07,sm07},
a proper determination of the QLF at $z\sim6$ is needed to constrain
the the quasar contribution.

In this paper we present the discovery of five $z\sim6$ quasars with 
$20<z_{AB}<21$ selected from 260 deg$^2$ of the SDSS southern survey, a deep 
imaging survey obtained by repeatedly scanning a 300 deg$^2$ area in the Fall 
Celestial Equatorial Stripe \citep{ade07a}. One of the five quasars, SDSS 
J020332.39+001229.3 (hereafter SDSS J0203+0012), was independently discovered 
by matching the UKIRT Infrared Deep Sky Survey \citep[UKIDSS;][]{war07} data 
to the SDSS data \citep{ven07}. These five quasars, together with another 
quasar, SDSS J000552.34--000655.8 (hereafter SDSS J0005--0006) previously
discovered in this region \citep{fan04}, form a well-defined low-luminosity 
quasar sample at high redshift. We use this sample and the luminous SDSS 
quasar sample to measure the QLF and constrain the quasar contribution to the 
reionization of the universe at $z\sim6$.

The structure of the paper is as follows. In $\S$ 2 we introduce the quasar
selection criteria and photometric and spectroscopic observations of quasar 
candidates. In $\S$ 3 we describe the properties of the five new
quasars. We derive the QLF at $z\sim6$ in $\S$ 4,
and discuss the contribution of quasars to the ionizing background in $\S$ 5. 
We give a brief summary in $\S$ 6. 
Throughout the paper we use a $\Lambda$-dominated flat
cosmology with H$_0=70$ km s$^{-1}$ Mpc$^{-1}$, $\Omega_{m}=0.3$, and
$\Omega_{\Lambda}=0.7$ \citep{spe07}.

\section{CANDIDATE SELECTION AND OBSERVATION}

\subsection{SDSS Deep Imaging Data}

The SDSS is an imaging and spectroscopic survey of the sky \citep{yor00} using 
a dedicated wide-field 2.5 m telescope \citep{gun06} at Apache Point 
Observatory. Imaging is carried out in drift-scan mode using a 142 mega-pixel 
camera \citep{gun98} which gathers data in five broad bands, $ugriz$, spanning 
the range from 3000 to 10,000 \AA\ \citep{fuk96}, on moonless photometric 
\citep{hog01} nights of good seeing. The effective exposure time is 54
seconds. The images are processed using specialized software \citep{lup01}, 
and are photometrically \citep{tuc06,ive04} and astrometrically 
\citep{pie03} calibrated using observations of a set of primary standard stars 
\citep{smi02} on a neighboring 20-inch telescope. All magnitudes are roughly 
on an AB system \citep{oke83}, and use the asinh scale described by 
\citet{lup99}. 

A primary goal of the SDSS imaging survey is to scan 8500 deg$^2$ of the north 
Galactic cap (hereafter referred to as the SDSS main survey). In addition to 
the main survey, SDSS also conducts a deep survey by repeatedly imaging a 300 
deg$^2$ area on the Celestial Equator in the south Galactic cap in the Fall
\citep[hereafter referred to as the SDSS deep survey;][]{ade07a}. This deep 
stripe (also called Stripe 82) spans $\rm 20^h<RA<4^h$ and 
$\rm -1.25\degr<Dec<1.25\degr$. The multi-epoch images, when coadded, allow 
the selection of much fainter quasars than the SDSS main survey. 
\citet{jia06b} have used the deep data to find low-redshift faint quasars 
selected from the SDSS coadded catalog, i.e., each run goes through the 
photometric pipeline PHOTO separately, and the resulting catalogs are coadded. 
The SDSS deep survey will eventually reach a depth of $i_{AB}=24.0\sim24.5$ 
(5 $\sigma$ detection for point sources) with 
more than 50 epochs of data \citep{ade07b}. The area and the depth will then 
be comparable to the CFHQS, which has covered $\sim400$ deg$^2$ to a limit of 
$i_{AB}\sim24.5$ \citep{wil07}.

At the time the coadded images used in this paper were made, 2005, a given 
area of sky on Stripe 82 had been scanned 5--18 times under standard SDSS 
imaging conditions, and all available data were included in this version of 
the coadds. The construction of the coadds is summarized as follows. First,
each input image was calibrated to a standard SDSS zeropoint using the median 
of the SDSS photometric solutions for the runs covering the area. For each 
image we made an inverse variance image to serve as a weight map, and within 
the map we assigned near-zero values to pixels with the INTERP bit set in the 
fpMask files of PHOTO, which indicates that a bad pixel, bad column, or cosmic 
ray has been interpolated over.
After the images were sky subtracted using the PHOTO sky estimate, each 
image was mapped onto a uniform rectangular output astrometric grid using 
a modified version of the registration software SWARP \citep{ber02}. The main 
modification was to incorporate the known SDSS camera distortions into the 
astrometry. The weight maps were subjected to the same mapping. The mapping 
used the LANCZOS3 kernel for the images and weights, while the INTERP bit 
images were mapped using NEAREST before being used to set near-zero values in 
the weight map images. The mapped images were then coadded using a flux scaled 
inverse-variance weighted average. The resulting images had known photometric 
and astrometric calibrations. For precision photometry, the PSFs as a 
function of position on the images were required. We used an algorithm 
(implemented in PHOTO) that coadded the PSFs as a function of position known 
from each input image using the same weights as the images were averaged.
Once the photometry, astrometry, and PSF were known, we proceeded to run a 
version of FRAMES (the main portion of PHOTO) 
slightly modified to take effective gains and sky noises into account when 
calculating error estimates. The uncalibrated output of PHOTO was adjusted and
placed on the SDSS AB system using a slightly modifed version of the SDSS 
target selection software TARGET. The details of the construction of the 
coadds will be left for a forthcoming paper.

The resulting data were tested in three ways. First, object by object
comparisons were made against single-run SDSS data. Second, object by
object comparisons were made with the SDSS coadded catalogs produced
by suitably averaging all the catalog information from the individual
runs covering Stripe 82 (courtesy of R. Scranton and D. Johnston).
Third, statistical internal measurements, such as stellar color-color
diagrams, were made. These tests show that the images were properly
coadded and the depth of the coadds is close to what was expected from
standard error propagation.  Figure 1 compares the photometric errors
of PSF magnitudes for point sources in the SDSS main survey data and
the SDSS coadded data. The error estimates were produced by the SDSS
photometric pipeline PHOTO. From Figure 1 the photometric errors in
the coadded data are significantly smaller than those in the
single-run data.
\begin{figure}
\plotone{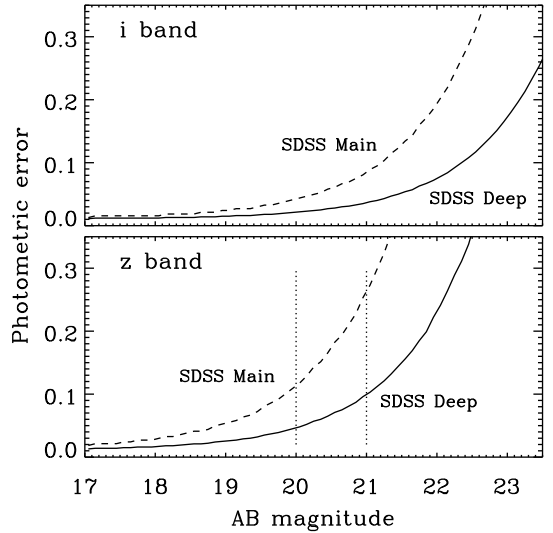}
\caption{Photometric errors of PSF magnitudes for point sources as a
  function of $i_{AB}$ and $z_{AB}$ magnitudes. The error estimates
  were produced by the SDSS photometric pipeline PHOTO. The errors in
  the SDSS coadded data are significantly smaller than those in the
  SDSS main survey data.  They are close to those expected from
  $\sigma_{\mathrm{single}}/N_{\mathrm{epoch}}^{1/2}$, where
    $\sigma_{\mathrm{single}}$ is the error in SDSS single-run data
    and $N_{\mathrm{epoch}}$ is the number of the runs used in the coadds. The
    flux limits of the $z\sim6$ quasar surveys used in the SDSS main
    survey \citep[e.g.][]{fan01a} and in our work are indicated as the
    two vertical dotted lines, at which $\sigma(z_{AB})\sim0.1$.}
\end{figure}

In this paper we used the data in the range $\rm 310\degr<RA<60\degr$,
as there were significantly fewer than 10 runs covering the range $\rm
300\degr<RA<310\degr$. The data also contains some ``holes'' in which
the coadded images were not available. The effective area for this
work is 260 deg$^2$. The median seeing as measured in the $riz$ bands
was $1.2\arcsec \pm 0.05\arcsec$, where the error is the standard
deviation of the seeing measured by PHOTO across the coadded images on
Stripe 82.

\subsection{Quasar Selection Procedure}

\begin{figure*}
\plotone{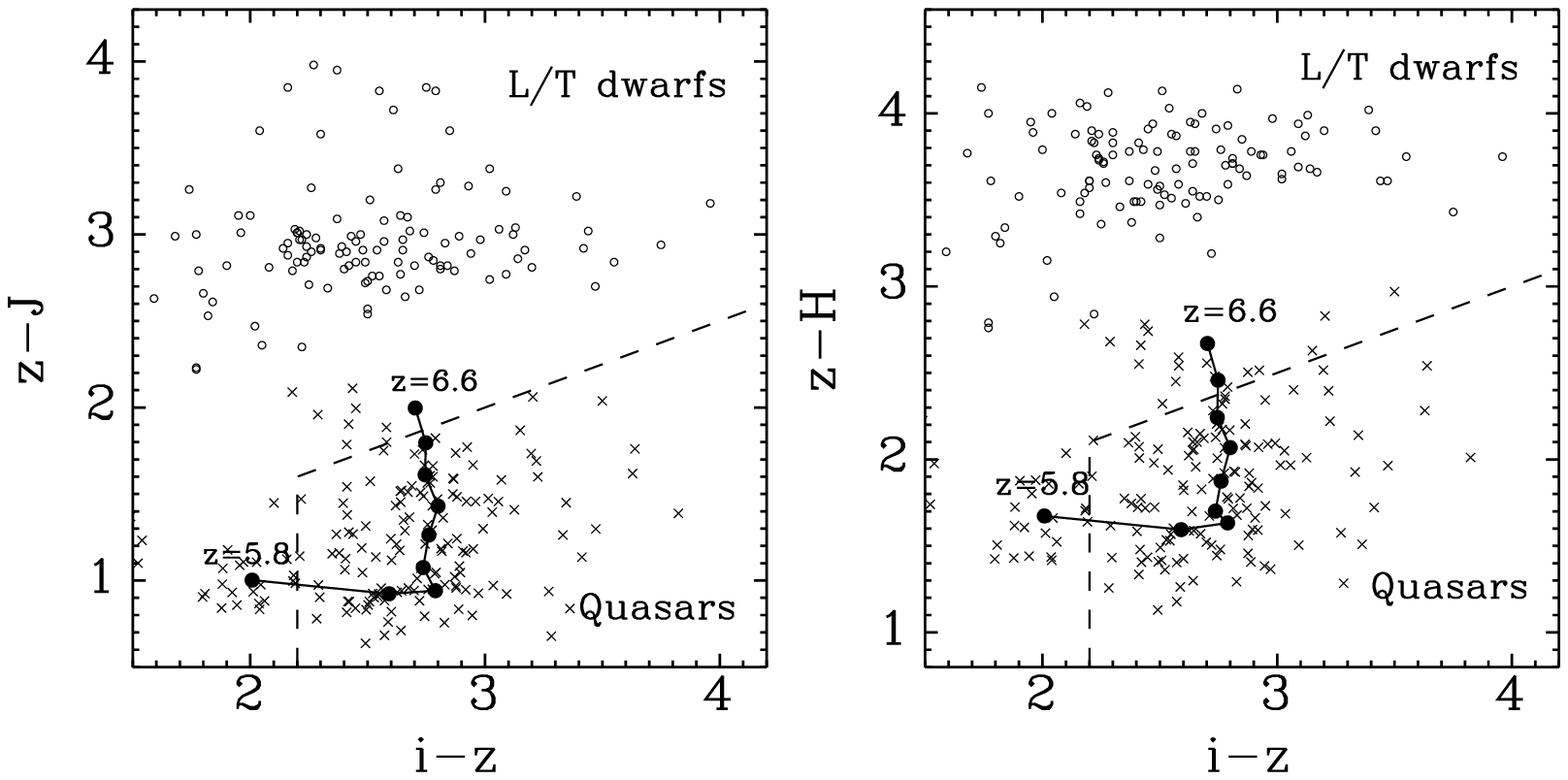}
\caption{The $i_{AB}-z_{AB}$ vs. $z_{AB}-J$ (and $H)$ color-color diagrams.
The open circles represent known L/T dwarfs from \citet{gol04}, \citet{kna04}, 
and \citet{chi06}. The crosses represent simulated quasars at $5.8<z<6.6$.
The median track of quasar colors as a function of redshift is shown as the 
filled circles.
We use the dashed lines to separate high-redshift quasar candidates 
from L/T dwarfs. Note that in practice there is no unambiguous separation 
between the dwarf locus and the quasar locus due to photometric errors.}
\end{figure*}
Because of the rarity of high-redshift quasars and overwhelming number of
contaminants, our selection procedure of $z>5.7$ faint quasars from the 
multi-epoch SDSS imaging data contains the following separate steps 
\citep[see also][]{fan01a,fan03}.
\begin{enumerate}
\item Select $i$-dropout sources from the SDSS deep stripe. 
Objects with $i_{AB}-z_{AB}>2.2$ and $z_{AB}<21$ that were not detected in
the $ugr$ bands were selected as $i$-dropout objects. We rejected sources
with one or more of the following SDSS processing flags: BRIGHT, EDGE, 
BLENDED, SATUR, MAYBE\_CR, and MAYBE\_EGHOST \citep[see][]{sto02}.
At $z>5.7$, the \lya\ emission line begins
to move out of the SDSS $i$ filter, so a simple cut of $i_{AB}-z_{AB}>2.2$ is 
used to separate high-redshift quasars (and cool brown dwarfs) from the
majority of stellar objects \citep[e.g.][]{fan01a}. 
At $z_{AB}=21$, the photometric errors of the coadded data reach 
$\sigma(z_{AB})\sim0.1$ as shown in Figure 1, so the $z_{AB}<21$ criterion 
guarantees a high quasar selection efficiency due to small photometric errors.
\item Remove false $i$-dropout objects. All $i$-dropout objects were visually 
inspected, and false detections were deleted from the list of candidates. The 
majority of the contaminants are cosmic rays. Although the SDSS photometric 
pipeline effectively rejects cosmic rays, the leakage of a tiny fraction of 
cosmic rays will contribute a large contamination to our sample. Cosmic rays
were recognized by comparing the individual multi-epoch images making up the
coadds. Brown dwarfs with high proper 
motions can be removed in a similar way, since the multi-epoch images were 
taken over a period of five years. In the selection of luminous $z\sim6$ 
quasars in the SDSS main survey, \citet{fan01a} used an additional step,
$z$-band photometry of $i$-dropout objects, to eliminate cosmic rays and
improve the photometry of potential candidates. In this work we did not
use this step, as the photometry of the coadds is robust. About 60 $i$-dropout
objects remained in this step.
\item Near-infrared (NIR) photometry of $i$-dropout objects. We then carried 
out NIR ($J$ or $H$ band) photometry of $i$-dropout objects selected from 
the previous step. The details of the NIR observations are described in $\S$ 
2.3. Using the $i_{AB}-z_{AB}$ vs. $z_{AB}-J$ (or $H)$ color-color diagrams
(Figure 2), high-redshift quasar candidates were separated from brown dwarfs 
(L and T dwarfs), which have more than ten times higher surface density. 
The open circles in Figure 2 represent known L/T dwarfs from \citet{gol04}, 
\citet{kna04}, and \citet{chi06}. The crosses represent simulated quasars at 
$5.8<z<6.6$. Although there is no clear separation between the dwarf locus and 
the quasar locus due to photometric errors, we selected quasars with the 
following criteria,
\begin{equation}
i_{AB}-z_{AB}>2.2\ \&\& \ z_{AB}-J<0.5(i_{AB}-z_{AB})+0.5,
\end{equation}
\begin{equation}
{\rm or}\ i_{AB}-z_{AB}>2.2\ \&\& \ z_{AB}-H<0.5(i_{AB}-z_{AB})+1.0.
\end{equation}
A total of 14 objects satisfied the criteria.
\item Follow-up spectroscopy of quasar candidates. The final step is to carry 
out optical spectroscopic observations of quasar candidates to identify
high-redshift quasars. The details of the spectroscopic observations
are described in $\S$ 2.3.
\end{enumerate}

\subsection{NIR Photometry and Optical Spectroscopic Observations}

\begin{figure*}
\plotone{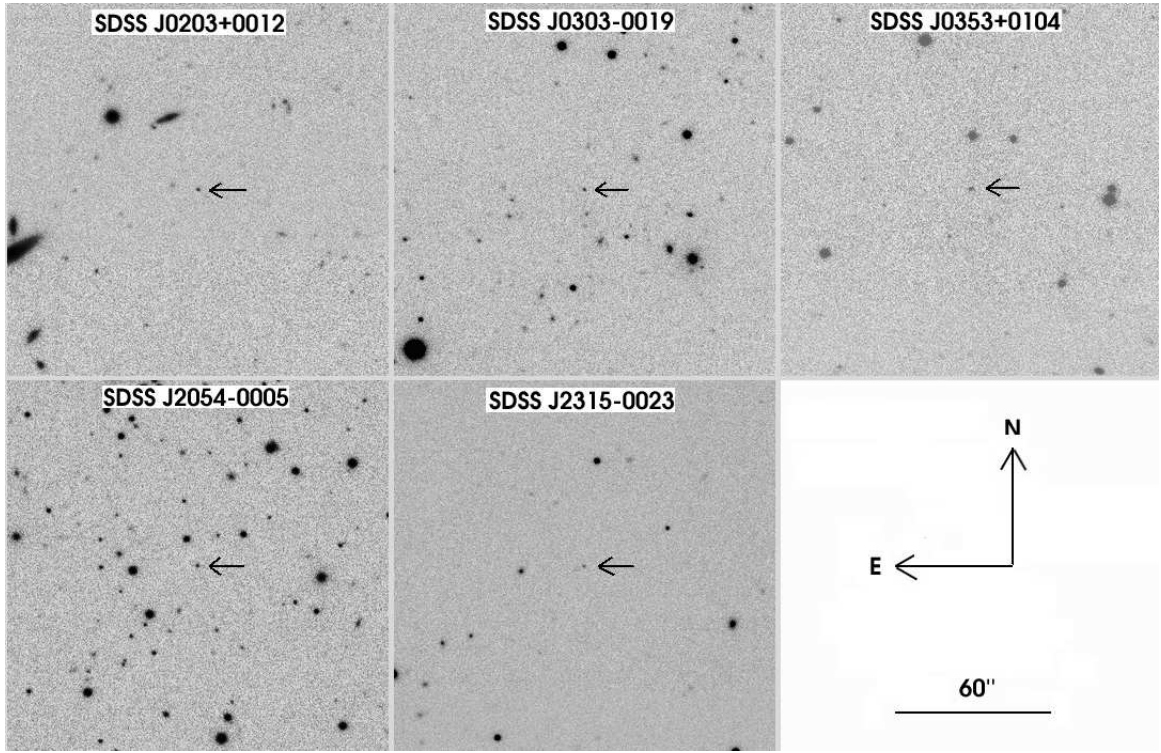}
\caption{The $z$-band finding charts of the five new $z\sim6$ quasars 
discovered in the SDSS deep stripe.}
\end{figure*}
\begin{deluxetable*}{cccccc}
\tablecaption{Optical and NIR Photometry}
\tablewidth{0pt}
\tablehead{\colhead{Quasar (SDSS)} & \colhead{Redshift\tablenotemark{a}} & 
  \colhead{$i_{AB}$ (mag)} &  \colhead{$z_{AB}$ (mag)} & 
  \colhead{$J$ (mag)} & \colhead{$H$ (mag)} }
\startdata
J000552.34$-$000655.8\tablenotemark{b} & 5.850$\pm$0.003 & $23.40\pm0.34$ 
  & $20.54\pm0.10$ & $19.87\pm0.10$ & $\ldots$        \\
J020332.39+001229.3\tablenotemark{c}   & 5.854$\pm$0.002 & $23.72\pm0.22$ 
  & $20.87\pm0.10$ & $19.05\pm0.08$ & $\ldots$        \\
J030331.40$-$001912.9                  & 6.070$\pm$0.001 & $23.92\pm0.23$ 
  & $20.85\pm0.07$ & $\ldots$       & $19.46\pm0.10$  \\
J035349.72+010404.4                    & 6.049$\pm$0.004 & $24.03\pm0.30$
  & $20.54\pm0.08$ & $\ldots$       & $18.55\pm0.06$  \\
J205406.49$-$000514.8                  & 6.062$\pm$0.004 & $23.30\pm0.22$
  & $20.72\pm0.09$ & $19.18\pm0.06$ & $\ldots$        \\
J231546.57$-$002358.1                  & 6.117$\pm$0.006 & $24.90\pm0.28$
  & $20.88\pm0.08$ & $19.94\pm0.08$ & $\ldots$        \\
\enddata
\tablenotetext{a}{The errors of the redshifts are the uncertainties obtained 
  from our fitting process.}
\tablenotetext{b}{This quasar was discovered by \citet{fan04}. The magnitudes 
  were taken from \citet{fan04}, and the redshift was determined from
  the \mgii\ emission line by \citet{kur07}.}
\tablenotetext{c}{This quasar was independently discovered by \citet{ven07}.}
\tablecomments{The $i_{AB}$ and $z_{AB}$ magnitudes are AB magnitudes and the 
  $J$ and $H$ magnitudes are Vega-based magnitudes.}
\end{deluxetable*}
In the first two steps we described above, we selected about 60 $i$-dropout 
objects with $i_{AB}-z_{AB}>2.2$ and $z_{AB}<21$ from 260 deg$^2$ of the 
SDSS coadded imaging data. After the selection of $i$-dropouts, we carried 
out $J$ or $H$-band\footnote{The $J$-band photometry is more efficient for
selecting $z\sim6$ quasars. Due to an instrument problem, however, the filter 
wheel was stuck and only the $H$ filter was available on the night of October 
2006.} photometry of these $i$-dropouts using the SAO Widefield InfraRed 
Camera (SWIRC) on the MMT in November 2005 and October 2006 to separate quasar 
candidates and cool dwarfs. We used a $5\times5$ dither pattern to obtain good 
sky subtraction and to remove cosmic rays. The exposure time at each dither 
position was 30 seconds. The total exposure time for each target was 
calculated to achieve an uncertainty of $\sigma_J$ (or $\sigma_H)\sim0.08$.
The typical exposure time on individual targets is 10 minutes.
The SWIRC data were reduced using standard IRAF\footnote{IRAF
is distributed by the National Optical Astronomy Observatories, which are
operated by the Association of Universities for Research in Astronomy, Inc.,
under cooperative agreement with the National Science Foundation.} routines.
Five bright objects from the Two Micron All Sky Survey \citep[2MASS;][]{skr06}
in the field of each target were used to apply the aperture correction and 
absolute flux calibration.

After the NIR photometry of the $i$-dropouts, we selected quasar candidates
that satisfied the criteria of Equation 1 or 2. Optical spectroscopy 
of the candidates was carried out using the Echelle Spectrograph and Imager 
\citep[ESI;][]{she02} on the Keck-II in January 2006 and the Low Dispersion 
Survey Spectrograph (LDSS-3) on Magellan-II in October 2006. 
The observations on ESI were performed in echellette mode, which provides 
excellent sensitivity from 4000 to 10,000 \AA. The observations on LDSS-3 were 
performed in longslit mode. LDSS-3 was designed to be very red sensitive. We 
used the VPH red grism with a ruling density of 660 lines/mm. The VPH red 
grism offers excellent throughput in the wavelength range from 6000 to 10,000 
\AA. The exposure time for each target was 20 minutes, which is sufficient to 
identify $z\sim6$ quasars in our sample. If a target was identified as a 
quasar, several further exposures were taken to improve 
the spectral quality. The quasar data were reduced using standard routines. 
After bias subtraction, flat-fielding, and wavelength calibration were applied 
to the frames, one-dimensional spectra were extracted, and were flux 
calibrated using the spectra of spectroscopic standard stars. 

\section{DISCOVERY OF FIVE NEW QUASARS AT {\boldmath $z\sim6$}}

From the spectroscopic observations on Keck/ESI and Magellan/LDSS-3 we 
discovered five $z\sim6$ quasars in the SDSS deep stripe. Their $z$-band 
finding charts are shown in Figure 3. Note that SDSS J0203+0012 was
independently discovered by \citet{ven07}. Another $z\sim6$ quasar in this 
area, SDSS J0005--0006 discovered by \citet{fan04}, was recovered by our 
selection criteria. These six quasars comprise a complete flux-limited sample 
at $z_{AB}<21$. The optical and NIR properties of the quasars are given in 
Table 1. The $i_{AB}$ and $z_{AB}$ magnitudes of the newly discovered quasars
are taken from the SDSS deep imaging data, and their $J$ and $H$ magnitudes 
are obtained from our MMT/SWIRC observations. All the quasars have $z_{AB}$
magnitudes between 20 and 21. The surface density of $z\sim6$ quasars
with $z_{AB}<20$ is about 1/470 deg$^2$ \citep{fan06a}, so it is reasonable
to find no quasars with $z_{AB}<20$ in a 260 deg$^2$ area.

The optical spectra of the six 
quasars are shown in Figure 4. The spectrum of SDSS J0005--0006 was taken from
\citet{fan04}. The spectra of SDSS J035349.72+010404.4 (hereafter SDSS 
J0353+0104) and SDSS J231546.57--002358.1 (hereafter SDSS J2315--0023) were 
taken on Keck/ESI with a total exposure time of 60 minutes on each source. The 
spectra of the other three quasars were obtained with Magellan/LDSS-3, and the
total exposure time on each source was 100 minutes. Each spectrum shown in 
Figure 4 has been scaled to the corresponding $z_{AB}$ magnitude
given in Table 1, thereby placed on an absolute flux scale.
\begin{figure*}
\epsscale{0.8}
\plotone{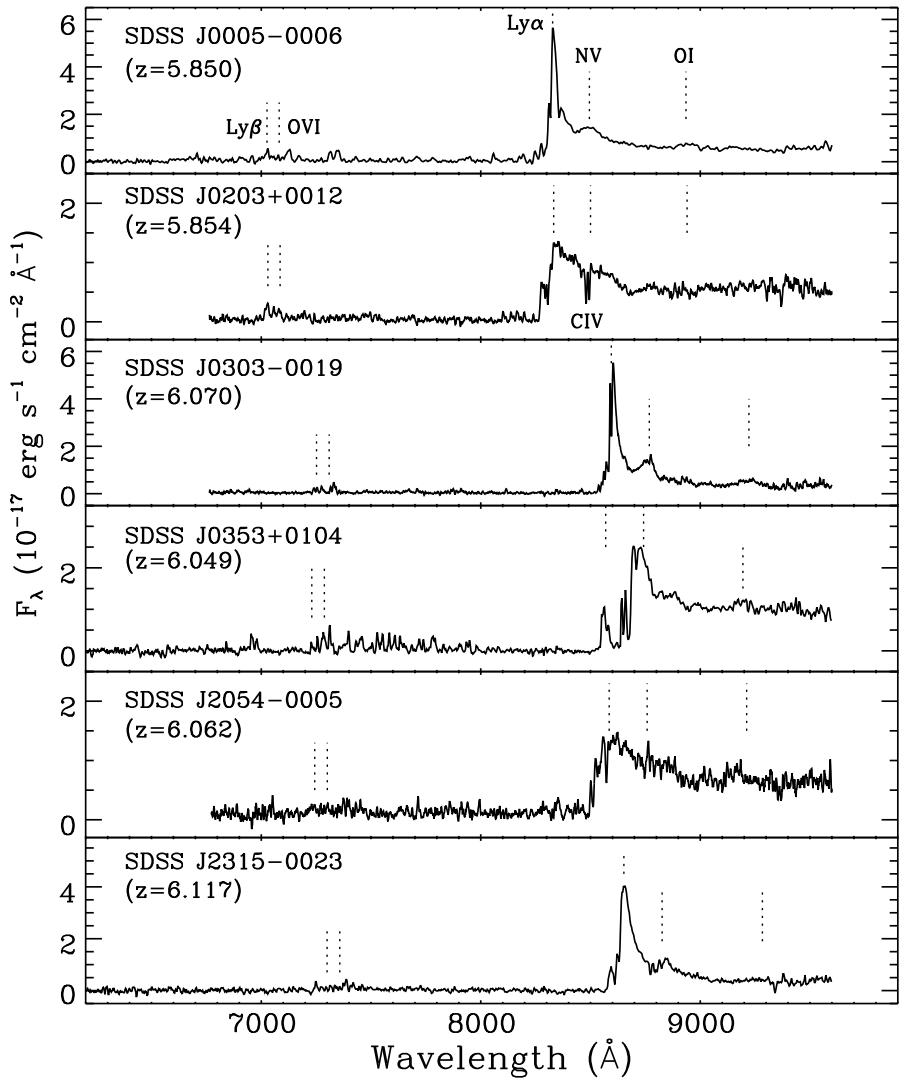}
\caption{Optical spectra of the six high-redshift quasars discovered in the 
SDSS deep stripe. The spectra of SDSS J0005$-$0006, SDSS J0353+0104, and
SDSS J2315$-$0023 were taken on Keck/ESI with a total exposure time of 60
minutes on each source. The spectra of the other three quasars were taken on
Magellan/LDSS-3 with a total exposure time of 100 minutes on each source.
The ESI spectra have been binned by 10 pixels and the LDSS-3 spectra been 
smoothed by 5 pixels. Four quasars in this sample are at 
$z>6$. Three quasars have narrow \lya\ emission lines. SDSS J0353+0104
is a BAL quasar.}
\end{figure*}

We estimate the redshifts for the new quasars from either the \lya, \nv\
$\lambda$1240 (hereafter \nv), or the \oi\ $\lambda1304$ (hereafter \oi) 
emission line. For each quasar, we measure the line center of one strong 
emission line using a Gaussian profile to fit the top $\sim50$\% of the line. 
This provides a rough estimate of the redshift. Using this redshift we 
subtract the power-law continuum and decompose the blended \lya\ and \nv\
emission lines into individual components. The details are described in the 
next paragraph. Then we determine the redshifts from individual emission 
lines. The redshift of SDSS J030331.40--001912.9 (hereafter SDSS J0303--0019) 
is measured from the \nv\ emission line, which is well separated from \lya\ 
due to the narrow line width. The measured redshift $6.070\pm0.001$ is 
consistent with the redshift $6.069\pm0.002$ determined from the weak \oi\ 
emission line. SDSS J0353+0104 is a BAL quasar as seen from strong 
absorption features around \lya, so its redshift is measured from the \oi\ 
emission line. 
The redshifts of the other three quasars are estimated from the \lya\ emission 
lines. They are usually biased because the blue side of \lya\ is affected by 
the \lya\ forest absorption \citep{sch91}. 
The mean shift with respect to the systemic 
redshift at $z>3$ is about 600 km s$^{-1}$ \citep{she07}, corresponding to 
$\delta z\sim0.015$ at $z\sim6$. 
We correct for this bias for the redshifts measured from \lya. The results are 
listed in Column 2 of Table 1. The errors in the table are the uncertainties 
obtained from our fitting process. For the redshifts measured from \lya,
their real errors could be much larger due to the scatter in the relation 
between \lya\ redshifts and systemic redshifts \citep{she07}.
In our sample four quasars have redshifts greater 
than 6. The most distant quasar, SDSS J2315--0023, is at $z=6.12$.

\begin{deluxetable*}{cccccc}
\tablecaption{Properties of the \lya\ and \nv\ Emission Lines}
\tablewidth{0pt}
\tablehead{\colhead{Quasar (SDSS)} & \colhead{Redshift} & \colhead{EW (\lya)} &
  \colhead{FWHM (\lya)} & \colhead{EW (\nv)} & \colhead{FWHM (\nv)}}
\startdata
J0005$-$0006 & 5.850 &  81.5$\pm$2.5 &  6.8$\pm$0.4 (1680 km s$^{-1}$)
  & 25.0$\pm$1.0 & 18.9$\pm$0.6 \\
J0203+0012   & 5.854 &  35.9$\pm$1.5 & 31.0$\pm$2.4 (7650 km s$^{-1}$)
  &  8.1$\pm$0.5 & 20.6$\pm$0.9 \\
J0303$-$0019 & 6.070 & 139.4$\pm$2.4 &  6.4$\pm$0.5 (1580 km s$^{-1}$)
  & 24.4$\pm$0.6 &  9.9$\pm$0.2 \\
J0353+0104\tablenotemark{a}& 6.049 & $\ldots$& $\ldots$& $\ldots$& $\ldots$ \\
J2054$-$0005 & 6.062 &  17.0$\pm$1.1 & 19.4$\pm$3.8 (4890 km s$^{-1}$)
  & 12.8$\pm$0.7 & 30.8$\pm$1.4 \\
J2315$-$0023 & 6.117 & 126.8$\pm$3.2 &  9.8$\pm$0.5 (2420 km s$^{-1}$)
  & 37.4$\pm$1.4 & 17.7$\pm$0.7 \\
\enddata
\tablenotetext{a}{This is a BAL quasar, so we did not measure its emission
line properties.}
\tablecomments{Rest-frame FWHM and EW are in units of \AA. The EW and FWHM
of the \lya\ emission lines have been corrected for \lya\ forest absorption.}
\end{deluxetable*}
We measure the rest-frame equivalent width (EW) and full width at half maximum
(FWHM) of \lya\ and \nv\ for each quasar except the BAL quasar SDSS 
J0353+0104. To allow the analysis of the emission lines we first fit and 
subtract the continuum. The wavelength coverage of each spectrum is too short 
to fit the continuum slope, so we assume it is a power law with a slope 
$\alpha_{\nu}=-0.5$ ($f_{\nu}\sim\nu^{\alpha_{\nu}}$), and normalize it to the 
spectrum at rest frame 1275--1295 \AA, a continuum window with little 
contribution from line emission. \lya\ and \nv\ are usually blended with each 
other, so we use three Gaussian profiles to simultaneously fit the 
two lines, with the first two profiles representing broad and narrow 
components of \lya\ and the third representing \nv. Since the blue side of the 
\lya\ emission line is strongly absorbed by \lya\ forest absorption systems, 
we only fit the red side of the line and assume that the line is symmetric. 
We ignore the weak \siii\ $\lambda1262$ emission line
on the red side of \nv. The measured EW and FWHM in units of \AA\
are shown in Table 2. We also give the FWHM of \lya\ in units of km s$^{-1}$.
We emphasize that the EW and FWHM of \lya\ in the table have taken into 
account the absorbed emission by the \lya\ forest, while most previous studies
did not take this absorption into account.

The distributions of the \lya\ EW and FWHM are broad. The average \lya\ EW and 
FWHM measured from the low-redshift SDSS composite spectrum of \citet{van01} 
are about 90 \AA\ and 20 \AA\ ($\sim5000$ km s$^{-1}$), respectively. 
The EW of \lya+\nv\ from a sample of quasars at $3.6<z<5.0$ is 69$\pm$18 \AA\
\citep{fan01b}, although this is affected by the \lya\ forest absorption.
We analyzed a sample of 20 luminous SDSS quasars at $z\sim6$, and find that
the mean \lya\ EW and FWHM are 56 \AA\ and 25 \AA\ (also corrected for the 
\lya\ forest absorption) with large scatters of 
40 \AA\ and 11 \AA, respectively.
In Table 2, three of our quasars (SDSS J0005--0006, SDSS J0303--0019, and SDSS 
J2315--0023) have \lya\ FWHM less than half of the typical value.
The \lya\ FWHM in both SDSS J0005--0006 and SDSS J0303--0019 is only 
$\sim1600$ km s$^{-1}$. However, their EW are close to or stronger than the 
typical \lya\ EW. On the contrary, the other two quasars, SDSS J0203+0012 
and SDSS J205406.49--000514.8 (hereafter SDSS J2054--0005), have typical \lya\ 
FWHM, but very weak \lya\ EW.

\begin{deluxetable}{cccc}
\tablecaption{Continuum Properties of the Quasars}
\tablewidth{0pt}
\tablehead{\colhead{Quasar (SDSS)} & \colhead{Redshift} & 
  \colhead{$m_{1450}$ (mag)} & \colhead{$M_{1450}$ (mag)}}
\startdata
J0005$-$0006 & 5.850 & 20.83$\pm$0.10 & --25.82$\pm$0.10 \\
J0203+0012   & 5.854 & 20.94$\pm$0.10 & --25.72$\pm$0.10 \\
J0303$-$0019 & 6.070 & 21.28$\pm$0.07 & --25.43$\pm$0.07 \\
J0353+0104   & 6.049 & 20.22$\pm$0.08 & --26.49$\pm$0.08 \\
J2054$-$0005 & 6.062 & 20.60$\pm$0.09 & --26.11$\pm$0.09 \\
J2315$-$0023 & 6.117 & 21.34$\pm$0.08 & --25.38$\pm$0.08 \\
\enddata
\end{deluxetable}
The best-fitting power-law continuum is also used to calculate $m_{1450}$ and
$M_{1450}$, the apparent and absolute AB magnitudes of the continuum at 
rest-frame 1450 \AA. The results are given in Table 3. The sample spans a 
luminosity range of $-26.5\le M_{1450}\le -25.4$. Because the \lya\ emission 
usually consists of a large fraction of the total emission in the $z$-band 
spectra of these quasars, the large scatter in the \lya\ EW results in a large 
scatter in the distributions of $m_{1450}$ and $M_{1450}$, even though the 
$z_{AB}$ magnitudes lie in the small range $20.5<z_{AB}<20.9$.

\subsection{Notes on individual objects}

{\bf SDSS J0005--0006 ($z=5.850$).} SDSS J0005--0006 was discovered by 
\citet{fan04}. This quasar has a very narrow \lya\ emission line. The 
rest-frame FWHM of \lya\ is only 1680 km s$^{-1}$ (after corrected for the 
\lya\ forest absorption). The strong 
\nv\ emission line is well separated from \lya. The central BH mass is
$3\times10^8\ M_{\sun}$ \citep{kur07}, an order of magnitude lower than the BH 
masses in luminous quasars at $z\sim6$
\citep[e.g.][]{bar03,ves04,jia07,kur07}. SDSS J0005--0006 was 
marginally detected in the $Spitzer$ IRAC 8.0$\mu$m band and was not detected 
in the $Spitzer$ MIPS 24$\mu$m band, indicating that there is no hot dust 
emission in this quasar \citep{jia06a}.

{\bf SDSS J0203+0012 ($z=5.854$).} SDSS J0203+0012 was independently 
discovered by matching the UKIDSS data to the SDSS data \citep{ven07}.
Its \lya\ emission line is broad but weak. The \lyb\ and \ovi\ $\lambda1033$ 
(hereafter \ovi) emission lines are clearly seen at $\sim7000$ \AA. A \civd\ 
absorption doublet is detected at $\lambda\lambda=8480.1$, 8494.3 \AA; this 
was also noticed by \citet{ven07}.

{\bf SDSS J0303--0019 ($z=6.070$).} SDSS J0303--0019 also has a very 
narrow \lya\ emission line. The rest-frame FWHM of \lya\ is 1580 km s$^{-1}$,
similar to the \lya\ width of SDSS J0005--0006. The strong \nv\ emission line 
is well separated from \lya.  The \lyb\ and \ovi\ emission lines are clearly 
detected at $\sim7300$ \AA. The continuum emission in this quasar is very
weak. The absolute magnitude $M_{1450}$ is $-25.43$, roughly two magnitudes
fainter than the luminous SDSS quasars at $z\sim6$.

{\bf SDSS J0353+0104 ($z=6.049$).} SDSS J0353+0104 is a BAL quasar, as seen 
from strong absorption features around the \lya\ emission line. Its redshift 
was measured from \oi. The fraction of BAL quasars in this small sample is one 
out of six, similar to the low-redshift fraction \citep{tru06}, although 
\civ\ observations may yield further BAL examples.
The \lyb\ and \ovi\ emission lines are seen at $\sim7300$ \AA.

{\bf SDSS J2054--0005 ($z=6.062$).} SDSS J2054--0005 has a very weak \lya\ 
emission line. The rest-frame EW of \lya\ is only 17.0 \AA, significantly 
smaller than the typical EW. But the FWHM of \lya, $\sim30$ \AA,
is similar to the mean value of \lya\ FWHM.

{\bf SDSS J2315--0023 ($z=6.117$).} SDSS J2315--0023 is the most
distant quasar in this sample.
The properties of the \lya\ and \nv\ emission lines are similar
to those of SDSS J0005--0006 and SDSS J0303--0019. It has a narrow but strong
\lya\ emission line. The rest-frame EW and FWHM of \lya\ are 127 \AA\ and
2420 km s$^{-1}$, respectively. It also has a very strong \nv\ emission line.

\section{QLF AT $z\sim6$}

The six quasars presented in this paper 
provide a flux-limited quasar sample at $z>5.8$. The survey
area is 260 deg$^2$ and the magnitude limit is $z_{AB}=21$. In this section
we calculate the spatial density of the $z>5.8$ quasars in the SDSS deep 
stripe, and combine this faint quasar sample with the SDSS bright quasar
sample to derive the QLF at $z\sim6$.

We use the selection function to correct the sample incompleteness due to
the selection criteria we applied. The selection function is defined as the 
probability that a quasar with a given magnitude, redshift, and intrinsic
spectral energy distribution (SED) meets our selection criteria. By assuming 
a distribution for the intrinsic SEDs, we calculate the 
average selection probability as a function of magnitude and redshift. To do 
this, we first calculate the synthetic distribution of quasar colors for a 
given ($M_{1450},z$), following the procedures in \citet{fan99} and 
\citet{fan01a}. Then we calculate the SDSS magnitudes from the model spectra 
and incorporate photometric errors into each band. For an object with given 
($M_{1450},z$), we generate a database of model quasars with the same 
($M_{1450},z$). The detection probability for this quasar is then the fraction 
of model quasars that meet the selection criteria. The details of the model
and simulation are described in \citet{fan99} and \citet{fan01a}.

\begin{figure}
\plotone{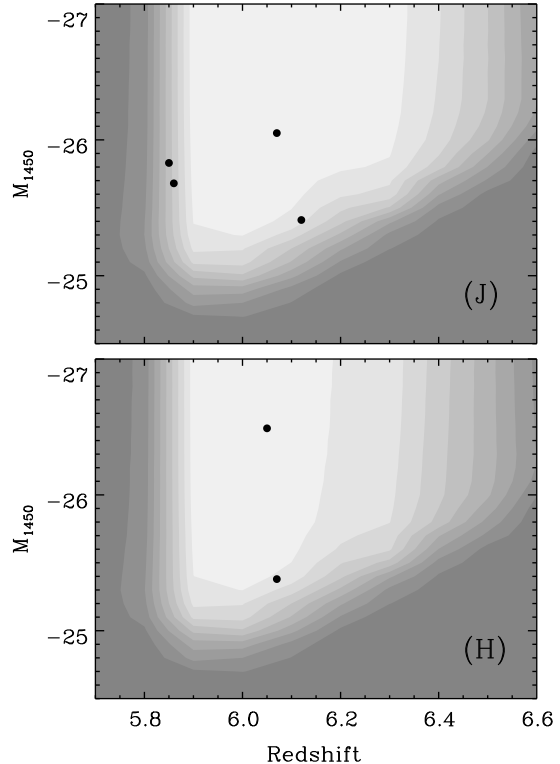}
\caption{Quasar selection function as a function of $M_{1450}$ and $z$ for the 
two selection criterion equations (1a and 1b) based on $J$ and $H$ bands. 
The contours in the 
figure are selection probabilities from 0.9 to 0.1 with an interval of 0.1.
The solid circles are the locations of the six $z\sim6$ quasars in our sample.}
\end{figure}
Figure 5 shows the selection function as a function of $M_{1450}$ and $z$
for the two selection criteria (Equations 1 and 2) based on $J$ and $H$ bands. 
The 
contours in the figure are selection probabilities from 0.9 to 0.1 with an
interval of 0.1.
The sharp decrease of the probability at $z\sim5.8$ is due to the color cut
of $i_{AB}-z_{AB}>2.2$. The two selection functions are slightly different.
Due to smaller photometric errors in the $i$ and $z$ bands, our survey probes 
$\sim1.5$ magnitude deeper than the SDSS main survey \citep[see][]{fan01a}.
The solid circles are the locations of the six $z\sim6$ quasars. All the 
quasars have the selection probabilities greater than 0.5.

We derive the spatial density of the $z>5.8$ quasars using the traditional 
$1/V_{a}$ method \citep{avn80}. The available volume for a quasar with 
absolute magnitude $M_{1450}$ and redshift $z$ in a magnitude bin
$\Delta M$ and a redshift bin $\Delta z$ is
\begin{equation}
V_{a} = \int_{\Delta M}\int_{\Delta z}p(M_{1450},z) \frac{dV}{dz} dz\,dM,
\end{equation}
where $p(M_{1450},z)$ is the selection function used to correct the sample 
incompleteness. We use one $M_{1450}$--$z$ bin for our small sample.
The redshift integral is over the redshift range $5.7<z<6.6$ and the magnitude
integral is over the range that the sample covers. The spatial density and its
statistical uncertainty can be written as
\begin{equation}
\rho = \sum_i \frac{1} {V_{a}^{i}}, \ \
\sigma(\rho) = \left[\sum_i \left(\frac{1} {V_{a}^{i}}\right)^2 \right]^{1/2},
\end{equation}
where the sum is over all quasars in the sample. This is similar to the 
revised $1/V_{a}$ method of \citet{pag00}, since $p(M_{1450},z)$ has already
corrected the incompleteness at the flux limit. We find that the spatial 
density at $\langle z\rangle=6.0$ and $\langle M_{1450}\rangle=-25.8$ is 
$\rho=(5.0\pm2.1)\times10^{-9}$ Mpc$^{-3}$ mag$^{-1}$.

\begin{figure}
\plotone{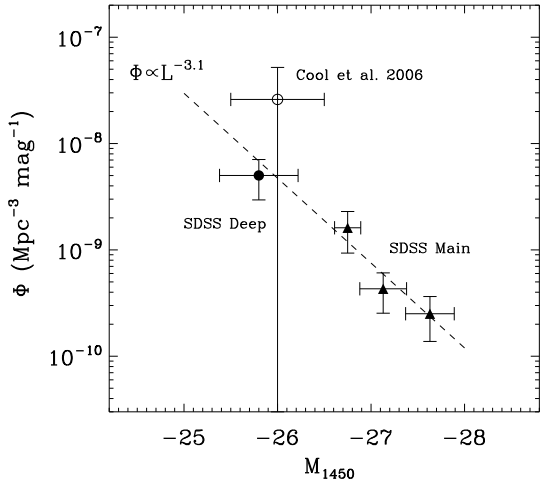}
\caption{QLF at $z\sim6$. The filled circle represents the density of the six 
quasars in the SDSS deep stripe and the filled triangles represent the 
densities from a study of 17 quasars from the SDSS main survey. 
The open circle is the constraint from
the quasar discovered by \citet{coo06}. The dashed line shows the best 
power-law fit to the SDSS quasars. The slope of the QLF is $-3.1\pm0.4$.}
\end{figure}
In the SDSS main survey, more than 20 quasars at $z>5.8$ have been discovered.
Seventeen of them were selected using similar criteria, and consist of a 
flux-limited quasar sample with $z_{AB}<20$. This bright quasar sample 
includes 14 published quasars \citep[e.g.][]{fan06a} and three new quasars 
(Fan et al. in preparation). We combine this sample with our sample to derive 
the QLF at $z\sim6$. The quasars in the combined sample are divided into four 
luminosity bins as shown in Figure 6. The QLF at the bright end at $z\sim6$ is 
well fit to a single power law 
$\Phi(L_{1450})\propto L_{1450}^{\beta}$, or,
\begin{equation}
\Phi(M_{1450})=\Phi^{\ast}10^{-0.4(\beta+1)(M_{1450}+26)},
\label{bslope}
\end{equation}
where we only consider luminosity dependence and neglect redshift evolution
over our narrow redshift range. The best fits are 
$\Phi^{\ast}=(5.2\pm1.9) \times 10^{-9}$ Mpc$^{-3}$ mag$^{-1}$ and 
$\beta=-3.1\pm0.4$.
The slope $\beta=-3.1\pm0.4$ is consistent with the slope $-3.2\pm0.7$ 
derived from the luminous sample alone by \citet{fan04} and with the slope
$>-4.63$ ($3\sigma$) constrained by the lack of lenses in four high-redshift 
quasars \citep{ric04}.

\section{DISCUSSION}

We have derived the QLF at $z\sim6$ and found a steep slope ($\beta=-3.1$) at
the bright end. From the SDSS Data Release Three, \citet{ric06} showed strong 
evidence that the bright-end slope of the QLF significantly flattens from 
$z\sim2$ to 4. They found that the slope at $z\sim2.0$ is $-3.1\pm0.1$ and 
the slope at $z\sim4.1$ flattens to $-2.1\pm0.1$. At $z\sim6$ the slope 
steepens again to $-3.1\pm0.4$ at a significance level of $\sim2.5\sigma$. 
The flattening of the slope at $z\sim4$ has been questioned by \citet{fon07},
who argued that the spectral index \citet{ric06} used to correct for sample
incompleteness was too blue. 
\citet{hop07} have analyzed the bolometric QLF from multiple surveys  
and stress that the flattening is seen at high significance. The lack  
of flattening claimed by \citet{fon07} would be real only if the  
distribution of quasar SEDs was redshift-dependent, contrary to what  
is found in most observations. Hence, the slope change from $z\sim4$  
to $z\sim6$ is highly likely to be physical.
The steepening of the slope at $z\sim6$ 
has important consequences in understanding early 
BH growth in quasars. Quasar evolution at $z\sim6$ is limited by the number of 
$e$-folding times available for BH accretion, therefore the shape of the QLF 
at $z\sim6$ puts strong constraints on models of BH growth
\citep[e.g.][]{wyi03,hop05,vol06,wyi06,li07}, and helps determine whether 
standard models of radiatively efficient Eddington accretion from stellar 
seeds are still allowed, or alternative models of BH birth (e.g. from 
intermediate-mass BHs) and BH accretion (super-Eddington or radiatively 
inefficient) are required \citep[e.g.][]{vol06}.

The steepening of the QLF slope also has a strong impact on the quasar
contribution to the ionizing background at $z\sim6$. The reionization
of the universe occurs at $z=11\pm4$ \citep{spe07} and ends at
$z\sim6$ \citep[e.g.][]{fan06b}. Studies have shown that quasars/AGN
alone are not likely to ionize the IGM at $z\sim6$
\citep[e.g.][]{dij04,mei05,wil05,dou07,sal07,srb07,sm07}. Galaxies
probably can provide enough photons for the reionization
\citep[e.g.][]{yan04,bou06,kas06,mcq07}, however, the individual
contributions of galaxies and quasars to the reionization are not well
determined. The galaxy contribution is uncertain due to our lack of
knowledge of factors such as the star-formation rate, the faint-end
slope of the galaxy luminosity function, and the escape fraction of
ionizing photons from galaxies \citep[e.g.][]{bun04,bou06}; while the
quasar contribution is poorly constrained due to the lack of the
knowledge of the faint end of the QLF at $z\sim6$.

We estimate the rate at which quasars emit ionizing photons at $z\sim6$ from 
the QLF derived in $\S$ 4. Following \citet{fan01a}, we calculate the photon 
emissivity of quasars per unit comoving volume at $z\sim6$ as
\begin{equation}
\dot{\cal N}_q=\epsilon_{1450}^q\,n_{1450}^p,
\end{equation}
where $\epsilon_{1450}^q$ is the quasar emissivity at 1450 \AA\ in units of
erg s$^{-1}$ Hz$^{-1}$ Mpc$^{-3}$ and $n_{1450}^p$ is the number of 
ionizing photons for a source with a luminosity of 1 erg s$^{-1}$ Hz$^{-1}$ at 
1450 \AA. We estimate $\epsilon_{1450}^q$ from
\begin{equation}
\epsilon_{1450}^q=\int \Phi(M_{1450})L_{1450}\,dM_{1450},
\end{equation}
where $\Phi(M_{1450})$ is the QLF at $z\sim6$ and the integral is over the
range $-30<M_{1450}<-16$. At low redshift, the shape of the QLF can be well 
modeled as a double power law with a steep bright end and a flat faint
end \citep[e.g.][]{boy00,ric05,jia06b}. At $z\sim6$, we assume a double 
power-law form for the QLF as well,
\begin{eqnarray}
&& \Phi(M_{1450})= \nonumber\\&&\frac{\Phi^{\ast}} 
  {10^{0.4(\alpha+1)(M_{1450}-M_{1450}^{\ast})}
  +10^{0.4(\beta+1)(M_{1450}-M_{1450}^{\ast})}},
\end{eqnarray}
where $\beta$ is the bright-end slope of the QLF, $\alpha$ is the faint-end
slope, and $M_{1450}^\ast$ corresponds to the characteristic luminosity
$L_{1450}^\ast$. At $z\leq2$,
$\alpha$ is roughly $-1.6$ and $M_{1450}^\ast$ is a function of redshift.
Little is known about $\alpha$ and $M_{1450}^\ast$ at $z>4$. At $z\sim6$, the
lower limit of $M_{1450}^\ast$ is about $-25$ as seen from Figure 6. Therefore 
we calculate $\epsilon_{1450}^q$ for a range of $\alpha$ between $-1.2$ and 
$-2.2$ and for a range of $M_{1450}^\ast$ between $-21$ and $-25$. The 
bright-end slope $\beta$ is fixed to $-3.1$ at the first step. We assume 
$\Phi^{\ast}$ to be the value derived from Equation~\ref{bslope}. This is a 
good approximation in the $\alpha$ and $M_{1450}^\ast$ ranges that we 
investigate here.

We calculate $n_{1450}^p$ by assuming the quasar SED following
\begin{equation}
L_{\nu} \propto \left\{ \begin{array}{ll}
                 \nu^{-0.5}, & \mbox{if $ \lambda >$ 1050 \AA;} \\
                 \nu^{-1.8}, & \mbox{if $ \lambda <$ 1050 \AA,}
                        \end{array}
 \right.
\end{equation}
and integrating the SED over an energy range of 1--4 Rydberg. The SED slope at
$\lambda>1050$ \AA\ is taken from \citet{van01} and at $\lambda<1050$ \AA\ 
from \citet{zhe97}.

The total photon emissivity per unit comoving volume required to ionize the 
universe is estimated to be
\begin{equation}
\dot{\cal N}_{ion}(z) = 10^{51.2}\, {\rm Mpc^{-3}\, s^{-1}} 
     \left({C\over 30}\right)\times \left(\frac{1+z}{6}\right)^3 
     \left({\Omega_b h^2\over 0.02}\right)^2
\label{nion}
\end{equation}
\citep{mad99}, where baryon density $\Omega_b h^2=0.02$ \citep{spe07} and $C$ 
is the clumping factor of the IGM. We examine the effects of three values 1, 
10, and 30 for $C$.

\begin{figure}
\plotone{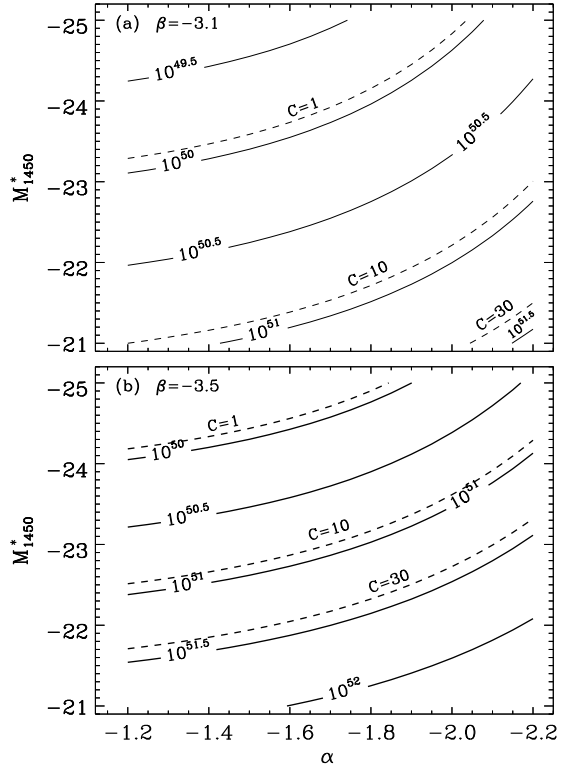}
\caption{Photon emissivity (per unit comoving volume) from quasars as a 
function of the faint-end QLF slope $\alpha$ and the characteristic 
luminosity $M_{1450}^\ast$. The solid lines represent the total photon
emissivity from quasars. The dashed lines show the photon emissivity required 
to ionize the IGM at $z\sim6$ for the clumping factors $C=1$, $C=10$, and 
$C=30$, respectively. The upper panel is for the bright-end QLF slope 
$\beta=-3.1$ and the lower panel is for $\beta=-3.5$.}
\end{figure}
The results are shown in Figure 7(a). The solid lines represent the photon 
emissivity per unit comoving volume from quasars as a function of the 
faint-end slope $\alpha$ and the characteristic luminosity $M_{1450}^\ast$.
The dashed lines show the photon emissivity required to ionize the IGM at 
$z\sim6$ for $C=1$, $C=10$, and $C=30$, respectively. 
There is an uncertainty of 0.4 ($1\sigma$) in the bright-end slope $\beta$, 
so we repeat our analysis for $\beta=-3.5$. The results are given 
in Figure 7(b). This figure shows that
\begin{enumerate}
\item The significance of the quasar contribution to the ionizing background
at $z\sim6$ depends on $\alpha$, $\beta$, $M_{1450}^\ast$, and $C$.
\item For a given $\beta$ (e.g. $\beta=-3.1$), the photon emissivity of 
quasars is a strong function of $M_{1450}^\ast$, but is not sensitive to 
$\alpha$ as long as $\alpha>-2.2$. Note that $\alpha$ is between 
$-1.0$ and $-2.0$ at low redshift.
\item In Equation~\ref{nion}, $\dot{\cal N}_{ion}(z)$ is a linear function
of the clumping factor $C$, which is critical here to determine whether or not 
quasars alone can ionize the universe. If $C=30$ \citep{gne97}, quasars can 
barely provide enough ionizing photons even if $M_{1450}^\ast\geq-22$. For a
homogeneous IGM ($C=1$), however, quasars can easily provide enough ionizing 
photons.
\item The photon emissivity from quasars increases significantly as $\beta$
steepens from $-3.1$ to $-3.5$. 
\end{enumerate}
Since the total photon emissivity of quasars is not sensitive to $\alpha$,
the quasar contribution to the reionization mainly depends on $M_{1450}^\ast$
and $C$ for a given $\beta$. It is clear that the quasar/AGN population can 
provide enough photons required to ionize the IGM only if the IGM is very 
homogeneous or the break luminosity is low.
At low redshift, $M_{1450}^\ast$ varies with 
redshift and the typical value is between --23 and --21. At $z\sim6$, the
lower limit of $M_{1450}^\ast$ is about $-25$.
To measure $M_{1450}^\ast$ at $z\sim6$, much deeper surveys are needed.
The clumping factor $C$ is also poorly constrained. Previous studies
used a range of values from 1 to $\sim$100. If the typical value of $C$ is
10--30 \citep[e.g.][]{gne97,mad99}, from Figure 7(a) the quasar population 
with $\beta=-3.1$ is not likely to provide enough ionizing photons.
However, if the reionization occurs outside-in and denser gas is ionized
at a later time when most of the volume of the universe has been reionized, 
clumpiness does not significantly increase the number of photons required for 
reionization \citep{mir00}. In this case, the equivalent $C$ is close to 1, 
and quasars can provide the required number of
photons to ionize the IGM.

\section{SUMMARY}

We have discovered five quasars at $z>5.8$ in 260 deg$^2$ of the SDSS deep 
stripe, including one previously discovered by \citet{ven07}. The most distant 
one is at $z=6.12$. These quasars were selected as $i$-dropout objects from 
the coadds of 10 SDSS imaging runs, going $\sim1.5$ magnitudes fainter than 
the SDSS main survey. The five quasars, with $20<z_{AB}<21$, are 1--2 
magnitudes fainter than the luminous 
$z\sim6$ quasars found in the SDSS main survey. The \lya\ emission lines in 
two quasars SDSS J0303--0019 and SDSS J2315--0023 are narrow 
($\rm FWHM\sim1600$ and 2400 km s$^{-1}$) but strong ($\rm EW\sim139$ and 127 
\AA), while the \lya\ emission lines in another two 
quasars SDSS J0303--0019 and SDSS J2315--0023 are broad ($\rm FWHM\sim7700$ 
and 4900 km s$^{-1}$) but weak ($\rm EW\sim36$ and 17 \AA). The 
fifth one, SDSS J0353+0104, is a BAL quasar.

The new quasars, together with a previously discovered quasar, SDSS 
J0005--0006, comprise a flux-limited quasar sample with $z_{AB}<21$ at 
$z\sim6$ over 260 deg$^2$. The sample covers the luminosity range
$-26.5\le M_{1450}\le -25.4$. The spatial density of the 
quasars at $\langle z\rangle=6.0$ and $\langle M_{1450}\rangle=-25.8$ is 
$(5.0\pm2.1)\times10^{-9}$ Mpc$^{-3}$ mag$^{-1}$. We use a single
power-law form to model the bright-end QLF at $z\sim6$ and find a slope 
of $-3.1\pm0.4$, which is significantly steeper than the slope of the QLF at
$z\sim4$. Using the derived QLF, we find that the quasar/AGN population can
provide enough photons required to ionize the IGM
at $z\sim6$ only if the IGM is very homogeneous and the characteristic
luminosity of the QLF is very low. To put better constraints on the quasar
contribution, much deeper surveys are needed.

The quasars in this paper were selected from the SDSS coadded images with 
5--18 runs.
Currently the SDSS deep stripe has been scanned between 40 and 50 times,
reaching $2\sim3$ magnitudes deeper than the main survey when co-added. We are 
performing a deeper survey of $z\sim6$ quasars down to $z_{AB}\sim22$ in 
this region. We expect to obtain a flux-limited sample with $z_{AB}<22$ in
the next few years.

\acknowledgments

We acknowledge support from NSF grant AST-0307384, a Sloan Research Fellowship
and a Packard Fellowship for Science and Engineering (LJ, XF). 
We thank the MMT staff, Magellan staff, and Keck staff for their expert help 
in preparing and carrying out the observations.

Funding for the SDSS and SDSS-II has been provided by the Alfred P. Sloan
Foundation, the Participating Institutions, the National Science Foundation,
the U.S. Department of Energy, the National Aeronautics and Space
Administration, the Japanese Monbukagakusho, the Max Planck Society, and the
Higher Education Funding Council for England.
The SDSS Web Site is http://www.sdss.org/.
The SDSS is managed by the Astrophysical Research Consortium for the
Participating Institutions. The Participating Institutions are the American
Museum of Natural History, Astrophysical Institute Potsdam, University of
Basel, Cambridge University, Case Western Reserve University, University of
Chicago, Drexel University, Fermilab, the Institute for Advanced Study, the
Japan Participation Group, Johns Hopkins University, the Joint Institute for
Nuclear Astrophysics, the Kavli Institute for Particle Astrophysics and
Cosmology, the Korean Scientist Group, the Chinese Academy of Sciences
(LAMOST), Los Alamos National Laboratory, the Max-Planck-Institute for
Astronomy (MPIA), the Max-Planck-Institute for Astrophysics (MPA), New Mexico
State University, Ohio State University, University of Pittsburgh, University
of Portsmouth, Princeton University, the United States Naval Observatory, and
the University of Washington.

\end{document}